\documentclass[amssymb,nobibnotes,superscriptaddress,longbibliography,twocolumn]{revtex4-1}

\usepackage{enumerate,amsmath}
\usepackage{graphicx}
\usepackage{color}

\newcommand{\mb}[1]{\mbox{\boldmath $#1$}}

\begin{document}

\title{Boosting computational power through spatial multiplexing \\
in quantum reservoir computing}%

\author{Kohei Nakajima}
\email{k\_nakajima@mech.t.u-tokyo.ac.jp}
\address{Chair for Frontier AI Education, Graduate School of Information Science and Technology, The University of Tokyo, Bunkyo-ku, 113-8656 Tokyo, Japan}
\address{JST, PRESTO, 4-1-8 Honcho, Kawaguchi, Saitama 332-0012, Japan}

\author{Keisuke Fujii}
\address{JST, PRESTO, 4-1-8 Honcho, Kawaguchi, Saitama 332-0012, Japan}
\address{The Hakubi Center for Advanced Research, Kyoto University, Yoshida-Ushinomiya-cho, Sakyo-ku, Kyoto 606-8302, Japan}
\address{Department of Physics, Graduate School of Science, Kyoto University, Kitashirakawa Oiwake-cho, Sakyo-ku, Kyoto 606-8502, Japan}

\author{Makoto Negoro}
\address{JST, PRESTO, 4-1-8 Honcho, Kawaguchi, Saitama 332-0012, Japan}
\address{Graduate School of Engineering Science, Osaka University, 1-3 Machikaneyama, Toyonaka, Osaka 560-8531, Japan}

\author{Kosuke Mitarai}
\address{Graduate School of Engineering Science, Osaka University, 1-3 Machikaneyama, Toyonaka, Osaka 560-8531, Japan}

\author{Masahiro Kitagawa}
\address{Graduate School of Engineering Science, Osaka University, 1-3 Machikaneyama, Toyonaka, Osaka 560-8531, Japan}

\date{\today}
\begin{abstract}
Quantum reservoir computing provides a framework for exploiting the natural dynamics of quantum systems as a computational resource. 
It can implement real-time signal processing and solve temporal machine learning problems in general, which requires memory and nonlinear mapping of the recent input stream using the quantum dynamics in computational supremacy region, where the classical simulation of the system is intractable.
A nuclear magnetic resonance spin-ensemble system is one of the realistic candidates for such physical implementations, which is currently available in laboratories. 
In this paper, considering these realistic experimental constraints for implementing the framework, we introduce a scheme, which we call a spatial multiplexing technique, to effectively boost the computational power of the platform. 
This technique exploits disjoint dynamics, which originate from multiple different quantum systems driven by common input streams in parallel. 
Accordingly, unlike designing a single large quantum system to increase the number of qubits for computational nodes, it is possible to prepare a huge number of qubits from multiple but small quantum systems, which are operationally easy to handle in laboratory experiments. 
We numerically demonstrate the effectiveness of the technique using several benchmark tasks and quantitatively investigate its specifications, range of validity, and limitations in detail.
\end{abstract}

\maketitle
\section{Introduction}
Recent developments in sensing and Internet of Things technology follow big data, which consists of a massive amount of complex time series data. 
Accordingly, a novel information processing technique that can deal with these data efficiently in real time is eagerly required. 
Conventional computers are, however, based on the von Neumann architecture, where the processor and memory are separately aligned, and this structure causes an intrinsic limitation in processing speed, which is called the von Neumann bottleneck.
Furthermore, the schemes of the von Neumann type models stipulate that to handle complex information processing, the computational system should be also built in a complex manner systematically.
While biological systems are complex systems that are constantly exposed to massive sensory data, they perform successful real-time information processing with lower computational costs and energy consumptions.
Their way of information processing is a typical non-von Neumann type, capitalizing on its natural and diverse dynamics, and has been a source of inspiration for many researchers \cite{IBM_Neuro}.

Reservoir computing is a framework for recurrent neural network training inspired by the way the brain processes information \cite{Jaeger0,Maass0,Reservoir}, and it provides a typical example of a non-von Neumann type computation \cite{Neuromorphic0}. 
A reservoir computing system consists of a high dimensional dynamical system, called a reservoir, driven by time-varying input streams, which generates transient dynamics with fading memory property and can perform nonlinear processing on inputs. 
Its framework can be used for real-time information processing with complex temporal structures, which makes it particularly suited to machine learning problems requiring memory, such as speech recognition, prediction of stock markets, and autonomous motor controls for robots. 
Conventionally, this scheme is implemented through randomly coupled artificial neural networks (i.e., echo state network (ESN) \cite{Jaeger0}) or through spiking neural networks (i.e., liquid state machine \cite{Maass0}) in the software program running on a PC.
As long as it runs on a conventional PC, the resulting computation is inevitably a von Neumann type.
On this basis, the physical implementations of the reservoir have been proposed to exploit the dynamics of native physics for information processing.
The implementations include the dynamics of the water surface \cite{Bucket}, photonics \cite{Laser0,Laser1}, spintronics \cite{Spintronics}, and the nanomaterials structured in the neuromorphic chip \cite{Neuromorphic0}.
Even the diverse body dynamics of soft robots have been shown to be used as a successful reservoir \cite{Kohei1,Kohei2,Kohei3,Kohei4}, suggesting that this framework could be applied to physical systems in various scales. 
Recently, quantum reservoir computing (QRC) has been proposed, which implements reservoir computing powered by quantum physics \cite{QR}.

Quantum dynamics is difficult to simulate using a conventional or classical computer due to the exponentially large degrees of freedom. 
This is generally termed a quantum computational supremacy, and the framework of QRC relies heavily on this property of quantum dynamics. 
Quantum reservoir (QR) dynamics are expressed as transitions of the basis states for quantum bits (qubits) driven by an input stream (Fig. \ref{fig1_new}A), which evolve over time through a unitary operator based on Hamiltonian. 
Signals are obtained through projective measurements from the system, called true nodes, which are used as direct reservoir states. 
An exponential number of degrees of freedom exist behind the measurement called hidden nodes, which affect the time evolution of the true nodes. 
The framework of QRC naturally takes into account the exponential degrees of freedom of quantum dynamics, which is intractable for the classical computer, for information processing.
Furthermore, the framework implements non-von Neumann type computing through a reservoir computing scheme, suggesting the full exploitation of assets from physical quantum dynamics.
It has been shown that the QR system can emulate nonlinear dynamical systems, including classical chaos, and exhibit robust information processing against noise \cite{QR}. 
As candidates for the physical experimental platform of the scheme, nuclear magnetic resonance (NMR)-spin ensemble systems \cite{NMRQC1,NMRQC2} have been proposed. 
In these systems, nuclear spins in molecules are used as the ensemble qubit system.
Usually, when monitoring a quantum system, its observables are affected by projective measurements, a process called backaction. 
In the NMR ensemble system, this effect of backaction can be neglected, and the signal can be successfully obtained by averaging the massive amount of copies of molecules existing in the ensemble system.

In this paper, we present a scheme to boost the computational power of QRs.
The most prominent and straightforward approach to improve the computational capability of the computational system is to increase its computational nodes.
In QRC, this primarily corresponds to increasing the number of qubits.
However, when viewed from a physical implementation standpoint (e.g., using an NMR spin-ensemble system), this approach requires a redesign or reconstruction of the sample molecules, which is operationally difficult and is energy and time consuming.

To overcome this problem, we introduce an effective approach to boost the computational power of the system using readily available small sample molecules, which are operationally easy to handle in the experiments.
Our scheme is called spatial multiplexing, in which we prepare multiple different small-sample molecules and inject common input streams to each system and use all the signals obtained from these systems as a big single reservoir system (Fig. \ref{fig1_new}B).
This procedure has previously been proposed in the applications of conventional ESNs, and many examples have demonstrated its effectiveness (e.g., \cite{Jaeger0}).
Here, we apply the scheme to QRC and present that its procedure is particularly suited to overcome the difficulty in a physically implemented reservoir setting.

In a software-implemented RC, since the scheme of spatial multiplexing exploits multiple disjoint ESNs as a new reservoir, it is operationally equivalent to assuming a single ESN having the same total number of computational nodes in the first place with a specific sparse internal weight matrix.
However, when viewing this scheme from physical RC perspectives, the situations are different.
In the NMR-implemented QRC, for example, even if the number of computational nodes are the same, the operational cost of preparing one huge sample molecule and that of preparing multiple small sample molecules are different.
By focusing on this operational difference, we can secure the scheme as one of the realistic and practical options to improve the computational power of physical reservoirs, which are often difficult to design freely and easily.
In the following sections, we argue the effectiveness of the spatial multiplexing technique for the NMR spin-ensemble system based QRC and quantitatively demonstrate how the scheme improves the computational performance in QRC. 
We also provide a detailed theoretical explanation of the specifications and range of validity of the scheme, which will be useful for evaluating other reservoir systems.

This paper is organized as follows. 
In the next section, we overview the formalization of QRC \cite{QR} and introduce spatial multiplexing into the setting. 
Subsequently, we theoretically examine the effect of spatial multiplexing in detail from a general standpoint. 
We then numerically demonstrate the power of the spatial multiplexing technique on QRC using conventional benchmark tasks in a machine learning context. 
Several approaches to engineer QRs through spatial multiplexing are also discussed. 
Finally, its practical aspect, future application domains in solving real world machine learning problems, and its implication to reservoir computing framework in general are discussed.

\section{Quantum reservoir computing through spatial multiplexing}
\subsection{Quantum reservoir dynamics}
\label{sec:QRC}
\begin{figure}
\centering
\includegraphics[width=80mm]{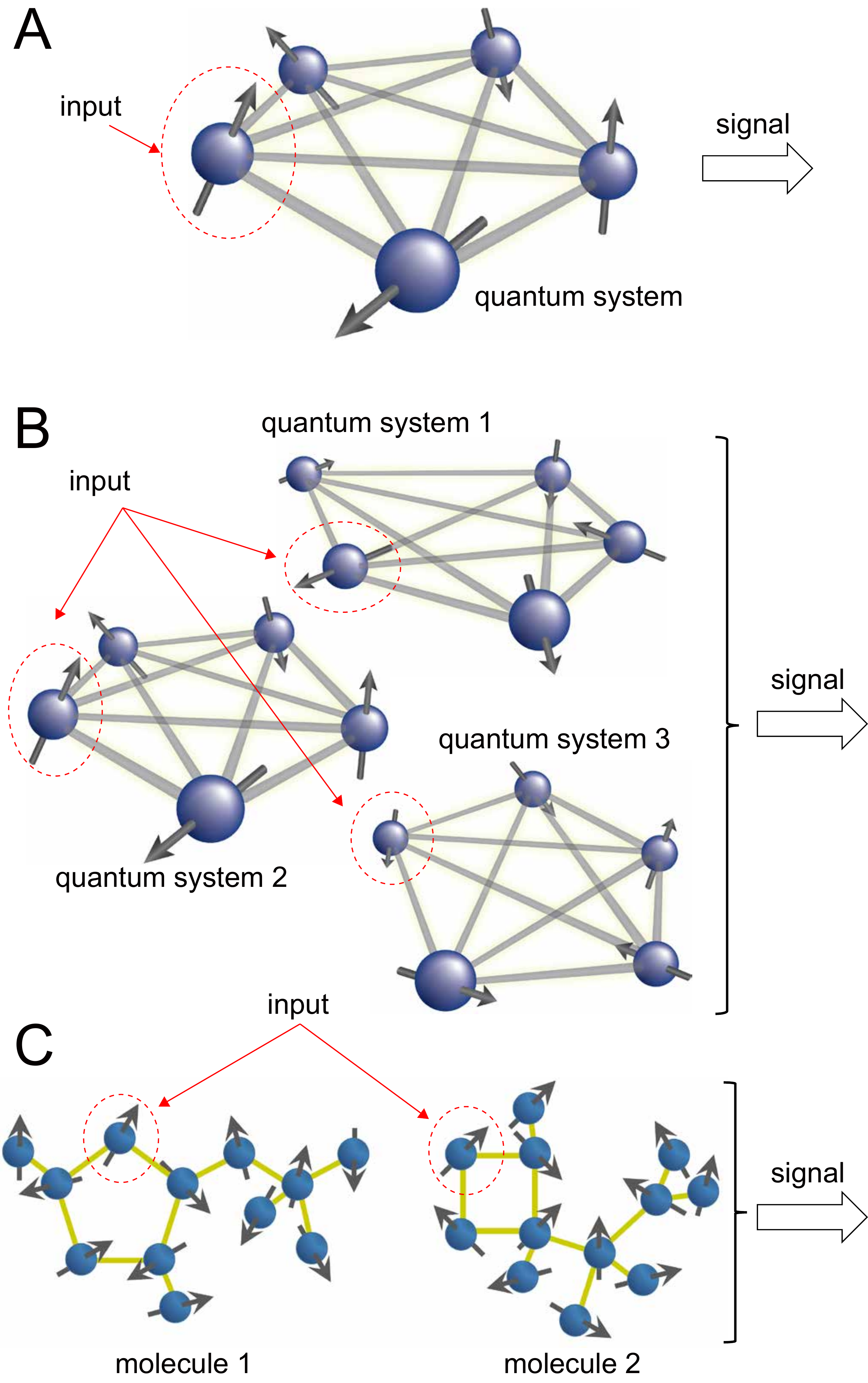}
\caption{{\bf Schematics explaining a spatial multiplexing technique for QRC.}
{\bf A.} A QRC system without spatial multiplexing for comparison, showing a quantum system with 5 qubits; the input is injected into the 1st qubit.
{\bf B.} A QRC system with spatial multiplexing. 
It shows multiple disjoint quantum systems, each containing 5 qubits; the input is injected into the 1st qubit for each system.
{\bf C.} Schematic of an experimental implementation of QRC with a spatial multiplexing technique for the NMR system.
}
\label{fig1_new}
\end{figure}
Let us consider a quantum state of an $N$-qubit system, which is described by a density operator $\rho$.
By denoting the Pauli operators to be 
\begin{eqnarray}
I=\sigma_{00}, X= \sigma _{10}, Z=\sigma _{01}, Y=\sigma _{11},
\end{eqnarray}
an $N$-qubit Pauli product is defined by $2N$-bit string $\mb{i}$:
\begin{eqnarray}
P(\mb{i}) =\bigotimes_{k=1}^{N} \sigma_{i_{2k-1}i_{2k}}.
\end{eqnarray}
By using the Pauli products $\{ P(\mb{i})\}$ as a basis of the operator space, the quantum state $\rho$ is represented by $4^N$ real vectors:
\begin{eqnarray}
\rho \rightarrow 
\mb{r}=\left(
\begin{array}{c}
r_{00...0}
\\
\vdots
\\
r_{11...1}
\end{array} 
\right)
\end{eqnarray}
where each element $r_{\mb{i}}$ is given in terms of the Schmidt-Hilbert inner product for the operator space as 
\begin{eqnarray}
r_{\mb{i}} = {\rm Tr}[ P(\mb{i}) \rho ]/2^{N}.
\end{eqnarray}
From the properties of the density operators,
\begin{eqnarray}
r_{00...0} = 1/2^{N},
-1 \leq r_{\mb{i}} \leq 1,
\sum _{\mb{i}} r^2_{\mb{i}} \leq 1.
\end{eqnarray}
In QRC, each element $r_{\mb{i}}$ is regarded as a {\it hidden} node of the network.
In quantum mechanics, any physical operation can be written as a linear transformation via a $4^N \times 4^N$ matrix $W$:
\begin{eqnarray}
\mb{r}' = W \mb{r}.
\label{eq:time_evolution}
\end{eqnarray}
The matrix $W$ can be constructed explicitly from the quantum operation $\mathcal{W}$ as follows:
\begin{eqnarray}
W_{\mb{j}\mb{i}} = {\rm Tr}\{ P(\mb{j})\mathcal{W}[P(\mb{i})] \} /2^N.
\end{eqnarray}
For example, a unitary dynamics under the Hamiltonian $H$ with a time interval ${\tau}$ is given by 
\begin{eqnarray}
(U_\tau)_{\mb{j}\mb{i}} = {\rm Tr}[ P(\mb{j}) e^{-i H \tau} P(\mb{i}) e^{i H \tau}] /2^N.
\end{eqnarray}

In order to exploit quantum dynamics for information processing, we have to introduce an input and the signals of the quantum system (see Fig.~\ref{fig1_new}A).
Suppose $\{ u_k \}$ is an input sequence, which is a continuous variable ($u_k \in [0,1]$).
(We consider the setting of one-dimensional input for simplicity, but its generalization to a multidimensional case is straightforward.)
A temporal learning task here is to find, using the quantum system, a nonlinear function $y_k=f(\{ u_l \}_{l=1}^{k})$ such that the mean square error between $y_k$ and a target output $\hat{y}_k$ for a given task becomes minimum.
Note that, as we see from Eq.~(\ref{eq:time_evolution}), there is no nonlinearity in each quantum operation $W$.
Instead, the time evolution $W$ can be changed according to the external input $u_k$, namely $W_{u_k}$, allowing the quantum reservoir to process the input information $\{u_k\}$ nonlinearly, by repetitively feeding the input.

Specifically, as an input, we replace the first qubit to the quantum state (Fig. \ref{fig1_new}A)
\begin{eqnarray}
\rho _{u_k} = \frac{I+(1-2 u_k)Z}{2}.
\end{eqnarray}
Corresponding matrix $S_{u_k}$ is given by 
\begin{eqnarray*}
(S_{u_k})_{\mb{j}\mb{i}} = {\rm Tr}\left\{ P(\mb{j}) \frac{I+(1-2u_k)Z}{2}  \otimes {\rm Tr}_{\rm 1st} [P(\mb{i})] \right\} /2^N,
\end{eqnarray*}
where ${\rm Tr}_{\rm 1st}$ indicates a partial trace with respect to the first qubit.
A unit time step is written as an input-depending linear transformation:
\begin{eqnarray}
\mb{r}((k+1)\tau) =  U_\tau S_{u_k}  \mb{r}(k\tau).
\end{eqnarray}
where $\mb{r}(k\tau)$ indicates the hidden nodes at time $k\tau$.

A set of observed nodes, which we call {\it true} nodes, $\{x_l\}_{l=1}^{M}$ is defined by a $4^N \times M$ matrix $R$,
\begin{eqnarray}
x_l (k\tau) = \sum _{\mb{i}} R_{l\mb{i}}r_{\mb{i}}(k\tau).
\end{eqnarray}
The number of true nodes $M$ has to be a polynomial in the number of qubits $N$.
That is, from exponentially many hidden nodes, a polynomial number of true nodes are obtained.
For simplicity, we take the single-qubit Pauli $Z$ operator on each qubit as the true nodes, i.e.,
\begin{eqnarray}
x_1 = r_{010...0}, x_2 = r_{00010...0}, ..., x_n = x_{0...01}.
\end{eqnarray}
Therefore, there is $M=N$ true nodes.
Figure \ref{fig2_new}A shows the typical reservoir dynamics driven by the input stream $\{ u_k \}$, and they consist of signals obtained from the true nodes.
Here we assume that the system is an ensemble quantum system, which consists of a huge number of copies of single quantum systems. 
Therefore, the signals from the true nodes are obtained without any backaction. 
Actually, the NMR spin ensemble system is such a system. 
A sample of an NMR spin-ensemble system contains typically $10^{18-20}$ copies of the same molecules. 
The magnetization of $10^{14-16}$ spins out of the sample can be measured with an RF coil with a sufficient SN ratio, while the remaining is not affected.

The unique feature of QRC in the reservoir computing context is that the exponentially many hidden nodes that originate from the dimensions of the Hilbert space 
are monitored from a polynomial number of signals defined as the true nodes.
Based on this setting, in the next section, two coordinated schemes are introduced to harness QR dynamics in a physically natural setting.
The first is called {\it temporal multiplexing}, which was already introduced in Ref. \cite{QR,TM_note}, and the second is called {\it spatial multiplexing}, which is a procedure applied to the QRC from this study.

\subsection{Temporal multiplexing}
In Ref.~\cite{QR}, temporal multiplexing has been found to be useful to extract complex dynamics on the exponentially large hidden nodes through the restricted number of true nodes.
In temporal multiplexing, the signals are sampled from the QR not only at the time $k \tau$, but also at each of the subdivided $V$ time intervals during the unitary evolution $U_{\tau}$ to construct $V$ virtual nodes, as shown in Fig.\ref{fig2_new}B (the upper diagram).
After each input by $S_{u_{k}}$, the signals are obtained for each subdivided intervals after the time evolution by $U_{ v \tau /V}$ ($v=1,2,...V$), i.e., 
\begin{eqnarray}
\mb{r}(k\tau +(v/V)\tau) \equiv U_{(v/V)\tau} S_{u_{k}} \mb{r}(k\tau).
\end{eqnarray}
Accordingly, as the QR system has $N$ true nodes, we have $NV$ corresponding computational nodes at each input timestep $k$ in total, and the virtual nodes are defined by
\begin{eqnarray}
x_{l}(k\tau +(v/V)\tau) =\sum _{\mb{i}} R_{l \mb{i}} r_{\mb{i}} (k\tau +(v/V)\tau).
\end{eqnarray}
This procedure allows us to make full use of input-driven transient dynamics, which can potentially include the influence of hidden nodes.
Using this technique, it is possible to effectively increase the total number of computational nodes employed in the learning process.
A similar technique can also be found, for example, in Ref.~\cite{Laser0} under the same motivations.

It is important to note that, as is obvious from the setting, the parameter $\tau$ modulates directly the dynamics of QR, while the parameter $V$ defines how we observe the dynamics. 
In Ref. \cite{QR}, the relevance of these parameters to the computational capability of the QR system is investigated.
It was observed that, according to the choice of the parameter $\tau$, the type of computation that can be performed well has changed, and the increase in the parameter $V$ essentially contributes to an improved computational performance.

\subsection{Spatial multiplexing}\label{sec:SM}
\begin{figure}
\centering
\includegraphics[width=85mm]{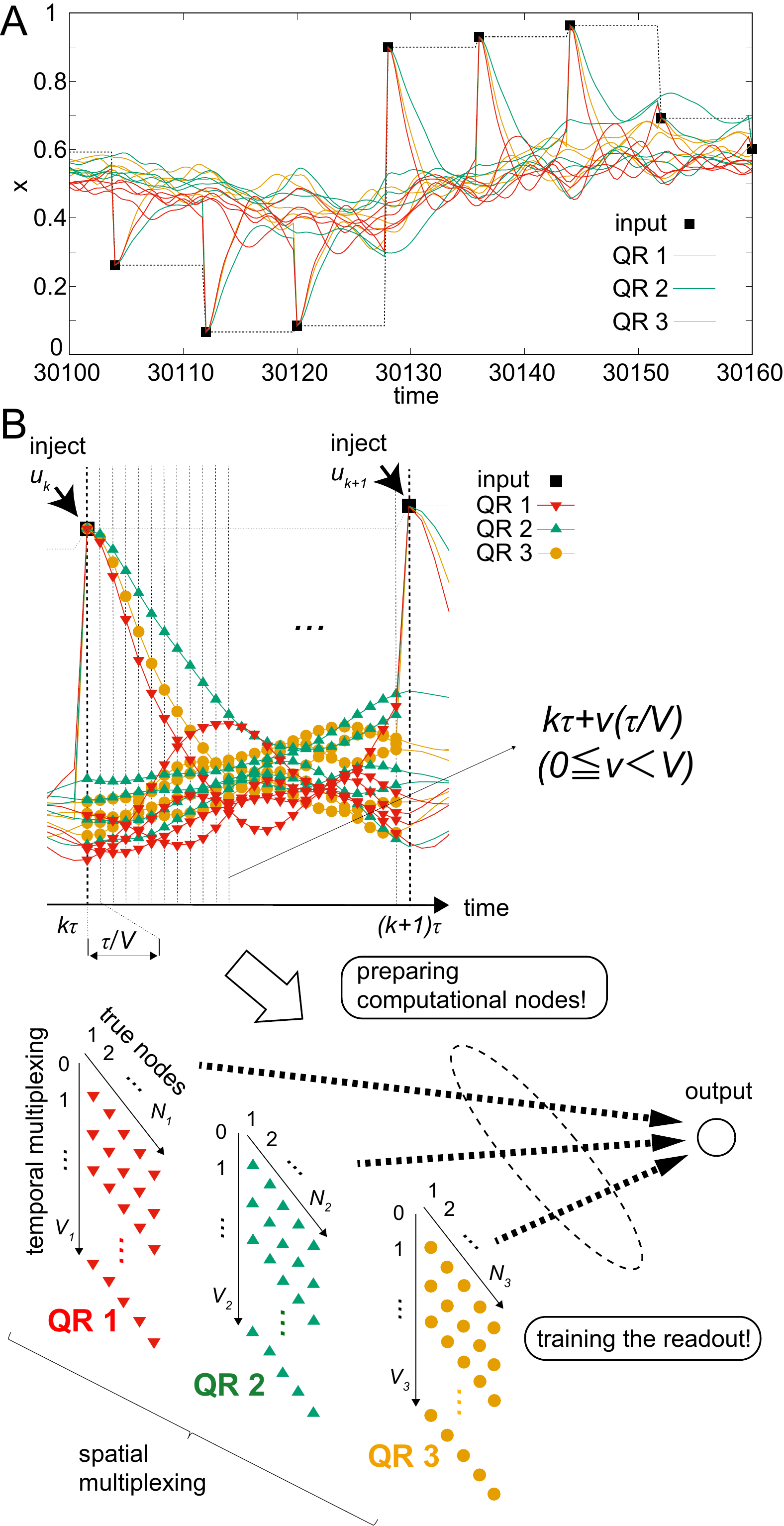}
\caption{{\bf Preparing computational nodes through spatial multiplexing.}
{\bf A.} A typical example of reservoir dynamics (time series of signals obtained from true nodes) in multiple disjoint quantum systems driven by a common input stream.
The plot overlays reservoir dynamics from three different quantum systems with the number of qubits set to 5 and $\tau \Delta =8$.
{\bf B.} Illustrating the temporal and spatial multiplexing scheme.
The upper diagram focuses on the time interval when the input $u_{k}$ is injected.
Signals from the three QR systems are overlaid, where the parameters $\tau_{c}, V_{c},$ and $N_{c}$ are set to be the same among the systems for simplicity.
The lower diagram expresses how to prepare the computational nodes in our settings.
The linear and static readout weight is attached to each computational node and the learning is performed by training the weights.
See text for details.
}
\label{fig2_new}
\end{figure}

Now, we consider boosting the computational power in QRC further.
The most straightforward and promising approach that comes to mind would be increasing the number of computational nodes.
This naturally leads to an increase in the number of qubits in the QR system.
(The approach of temporal multiplexing, which secures virtual nodes from the signals, is also reasonable in terms of increasing the number of computational nodes.)
Considering the physical implementations of QRC to the NMR system, however, as we explained previously, this procedure of increasing the number of qubits corresponds to the enlargement and redesign of sample molecules, and it is not always easy in practice.
In the NMR system, the local control and measurement of a qubit is accomplished with the difference of the resonant frequency.
The resonant frequency differs from the nuclear species.
Among many species, only few species such as $^{1}$H, $^{13}$C, $^{15}$N, and $^{19}$F are easy to handle and thus used as qubits before.
The resonant frequency is also slightly shifted due to the difference of the chemical environment even with the same species, which enable us the local control of them.
However, it is not easy to design and synthesize a molecule that includes many addressable spins with the different species and environment.
Since a 12 addressable spin system in a liquid has been developed in 2006 \cite{Negoro1}, the record still remains unbroken.

In this study, based on these physical constraints of the experimental settings, we introduce an effective and practical procedure to increase the computational resource, which is relatively easy to implement under the practical condition.
The procedure is called {\it spatial multiplexing}.
We prepare multiple disjoint QR systems, which are spatially distant or uncoupled, and we drive them with a common input stream in parallel (Fig. \ref{fig1_new}B).
Subsequently, we collect the signals from each QR system in the previously explained manner, and we use all of these signals from different QR systems as one entire set of reservoir dynamics.  
For the NMR-implemented QRC, this approach enables the exploitation of readily available sample molecules, which already exist in the laboratory, to increase the number of computational nodes. 
Compared to redesigning the sample molecules as a computational resource, this approach should be relatively handy and practical for experimenters.
For example, the aforementioned 12 qubit molecule and another one developed in 2017 \cite{Negoro2} can be potentially utilized for spatial multiplexed reservoirs (Fig. \ref{fig1_new}C).
To synthesize other 12 qubit molecules based on the developed molecules with chemical modifications may be easier than a molecule with more qubits.

Let us consider $C$ different QRs driven by a common input stream $u_{k}$ in parallel.
For each QR system $c$, which has $N_{c}$ qubits with a corresponding number of true nodes, the time interval to inject input $\tau_{c}$ and the corresponding unitary evolution $U_{\tau_{c}}$ can be set differently (Fig. \ref{fig2_new}B).
Each QR system is also equipped with temporal multiplexing having $V_{c}$ virtual nodes (Fig. \ref{fig2_new}B).
As a result, the spatial multiplexing induces $\sum_{c=1}^{C} N_{c} V_{c}$ nodes in total, and these computational nodes are exploited as a single reservoir.
We investigate systematically whether the procedure of spatial multiplexing really boosts the computational power of the QRC or not and, furthermore, to what extent it improves the performance in detail in the later sections.

\subsection{Output settings and learning procedure}\label{learning}
In the reservoir computing approach, the output is obtained as a weighted sum of the reservoir states, and the learning of a target function is executed by training linear and static readout weights attached to the reservoir nodes in a supervised manner.
Here, we explain how to train the readout weights from the observed signals of QR after the procedures of temporal and spatial multiplexing.
According to the previous sections, temporal and spatial multiplexing introduces $N_{total}=\sum_{c=1}^{C} N_{c} V_{c}$ computational nodes in total (Fig. \ref{fig2_new}B).
The state of the computational node $i$ at timestep $k$ is expressed as $x'_{ki}$ by rearranging the subscript from the original, and we introduce a constant bias term $x'_{k0} =1.0$.
The system output of the system is then expressed as
\begin{eqnarray}
y_k = \sum_{i=0}^{N_{total}}x'_{ki}  w_{i},
\end{eqnarray}
where $w_{i}$ is a linear and static weight attached to node $i$.
Let $\{\hat{y}_k\}_{k=1}^L$ be the target sequence for learning, where $L$ is the length of the training phase that is assumed much greater than $N_{total}+1$ and the training of the readout weights $\{ w_{i} \}_{i=0}^{N_{total}}$ is to minimize $\sum_{k=1}^{L}(y_k- \hat{y}_k)^2$.
By collecting the target output $\hat{y}=[\hat{y}_{1}, \hat{y}_{2}, ..., \hat{y}_{L}]^{\mathrm{T}} $ and the corresponding $N_{total}+1$ reservoir states in the learning phase as the training data matrix $X$, which is a $L \times (N_{total} +1)$ matrix, the optimal weight $\hat{w}=[\hat{w}_{0}, \hat{w}_{1}, ..., \hat{w}_{N_{total}}]^{\mathrm{T}}$ can be obtained as a least squares solution $\hat{w}= (X^{\mathrm{T}}X)^{-1}X^{\mathrm{T}}\hat{y}$.

As we see later in detail, when we actually let the QR system perform the computational tasks in this study, the experimental trial consists of a washout phase, training phase, and evaluation phase.
The washout phase is to eliminate the influence of initial transients of the reservoir states, and the trained readout weights in the training phase are exploited to generate outputs in the evaluation phase.

\subsection{Theoretical insights into the effect of spatial multiplexing}\label{theory}
In this section, we investigate theoretically the effect of spatial multiplexing, and we show its range of validity and limitations.
The argument in this section is not limited to quantum system but is generally applicable to any reservoir system.
Initially, we prove concisely that the procedure of spatial multiplexing always improves computational performance (or, at worst, will not change the performance).

Let us assume that we have two reservoirs, reservoirs A and B, which have $N_{A}$ and $N_{B}$ computational nodes, respectively.
Consider the corresponding regression equations, $y=X_{A}w_{A} + r_{A}$ and $y=X_{B}w_{B} + r_{B}$, where $X_{A}$ is a $T \times N_{A}$ matrix and $X_{B}$ is a $T \times N_{B}$ matrix with realizations $T$ satisfying $N_{A} + N_{B} \leq T$, and $r_{A}$ and $r_{B}$ are residuals.
We assume that $X_{A}$ and $X_{B}$ are full rank, and $w_{A}$ and $w_{B}$ are least squares solutions expressed as $w_{A} = (X_{A}^{\mathrm{T}} X_{A})^{-1} X_{A}^{\mathrm{T}} y$ and $w_{B} = (X_{B}^{\mathrm{T}} X_{B})^{-1} X_{B}^{\mathrm{T}} y$, respectively.
With projectors $P_{A} = X_{A}(X_{A}^{\mathrm{T}}X_{A})^{-1}X_{A}^{\mathrm{T}}$ and $P_{B} = X_{B}(X_{B}^{\mathrm{T}}X_{B})^{-1}X_{B}^{\mathrm{T}}$, $X_{A}w_{A}=P_{A}y$ and $X_{B}w_{B}=P_{B}y$.
Accordingly, the residuals can be expressed as $r_{A}^{2} = ||y- X_{A}w_{A}||^{2}$ = $||(I-P_{A})y||^{2}$ and $r_{B}^{2} = ||y- X_{B}w_{B}||^{2}$ = $||(I-P_{B})y||^{2}$.
We consider combining reservoirs A and B and constructing a new reservoir ``A+B.''
Similarly, for $X_{A+B} = \begin{pmatrix} X_{A} & X_{B} \end{pmatrix}$, $y=X_{A+B}w_{A+B} + r_{A+B}$, where $w_{A+B}$ is a least squares solution expressed as $w_{A+B} = (X_{A+B}^{\mathrm{T}} X_{A+B})^{-1} X_{A+B}^{\mathrm{T}} y$ and $r_{A+B}$ is a residual.
We assume that $X_{A+B}$ is full rank. 
With projectors $P_{A+B} = X_{A+B}(X_{A+B}^{\mathrm{T}}X_{A+B})^{-1}X_{A+B}^{\mathrm{T}}$, $X_{A+B}w_{A+B}=P_{A+B}y$, and a residual can be expressed as $r_{A+B}^{2} = ||y- X_{A+B}w_{A+B}||^{2}$ = $||(I-P_{A+B})y||^{2}$.
Because $w_{A+B}$ is a least squares solution,
\begin{align*}
r_{A+B}^{2} &= ||(I-P_{A+B})y||^{2} \\
&= ||y-X_{A+B}w_{A+B}||^{2} \\ 
&\leq ||y-X_{A+B}\begin{pmatrix}
w_{A} \\
0 \\
\end{pmatrix} ||^{2} \\
&= ||y- \begin{pmatrix} X_{A} & X_{B} \end{pmatrix} \begin{pmatrix}
w_{A} \\
0 \\
\end{pmatrix} ||^{2} \\
&= ||y-X_{A}w_{A}||^{2} \\ 
&= ||(I-P_{A})y||^{2} = r_{A}^{2}.
\end{align*}
The equal sign can be used only when $P_{A+B} = P_{A}$.
Likewise, $r_{A+B}^{2}\leq r_{B}^{2}$ and $r_{A+B}^{2}\leq \min\{ r_{A}^{2}, r_{B}^{2} \}$ holds.
Actually, this relation shows the reason why the performance improves by increasing the computational nodes in the system in general.
It also suggests that the couplings and interactions within the reservoir are not explicitly required for this improvement in theory.

Estimating the upper limit of the improvement in performance in terms of how the residual decreases by adding the reservoir B to the reservoir A is also possible.
In general, projector $P$ satisfies $P = P^{\mathrm{T}}$, $PP = P$, and if $P$ is a projector, then $I-P$ is also a projector.
Applying these properties to the above results, with a few transformations, we obtain
\begin{align*}
0 \leq r_{A}^{2} - r_{A+B}^{2} = \langle (P_{A+B}-P_{A})y, y \rangle,
\end{align*}
where $\langle \cdot, \cdot \rangle$ is an inner product.
This relation suggests that $Q_{A, A+B} := P_{A+B} - P_{A}$ is positive semidefinite, and the largest eigenvalue is
\begin{align*}
\lambda_{Q_{A, A+B}} := \sup_{y \neq 0} \frac{\langle (P_{A+B}-P_{A})y, y \rangle}{\langle y, y \rangle}.
\end{align*}
Thus,
\begin{align*}
0 \leq r_{A}^{2} - r_{A+B}^{2} \leq \lambda_{Q_{A, A+B}} ||y||^{2},
\end{align*}
where $\lambda_{Q_{A, A+B}}$ expresses the supremum of the reductions of a normalized residual when adding reservoir B to reservoir A.
Similarly, we obtain $0 \leq r_{B}^{2} - r_{A+B}^{2} \leq \lambda_{Q_{B, A+B}} ||y||^{2}$, where $Q_{B, A+B} := P_{A+B} - P_{B}$ is also positive semidefinite and $\lambda_{Q_{B, A+B}}$ is the largest eigenvalue of $Q_{B, A+B}$.
Using the above relations for $r_{B}^{2} - r_{A+B}^{2}$ and $r_{A}^{2} - r_{A+B}^{2}$, and $r_{A+B}^{2}\leq \min\{ r_{A}^{2}, r_{B}^{2} \}$, we obtain
\begin{align*}
&\max\{(r_{A}^{2}-\lambda_{Q_{A, A+B}} ||y||^{2}), (r_{B}^{2}-\lambda_{Q_{B, A+B}} ||y||^{2}) \} \\
&\leq r_{A+B}^{2} \leq \min\{ r_{A}^{2}, r_{B}^{2} \},
\end{align*}
where $0 \leq \max\{(r_{A}^{2}-\lambda_{Q_{A, A+B}} ||y||^{2}), (r_{B}^{2}-\lambda_{Q_{B, A+B}} ||y||^{2}) \}$.
Thus, we can evaluate and predict the extent of the improvement without actually performing the task using reservoir A+B.  

We should be careful because the above facts do not always hold in practice.
Two points should be noted.
The first is overfitting.
Spatial multiplexing can increase computational nodes drastically, so we should be careful when balancing between the size of training data set and the system size.
Since spatial multiplexing always results in an improved performance for training data set, if the performance worsens with spatial multiplexing in the evaluation phase, we can infer back that it is caused by overfitting.

Second, the above facts are based on the assumption that $X_{A}$, $X_{B}$, and $X_{A+B}$ are full rank.
This condition does not always hold in practice.
A typical example is a case in which synchronization occurs, which makes the reservoir dynamics identical or low-dimensional.
Notably, even if no coupling exists between the reservoirs in spatial multiplexing, the synchronization can still occur.
This is a phenomenon often called common input synchronization \cite{EdgeofChaos}, or when the driving input is a random signal, it is called common noise synchronization \cite{CommonNoise}.
Ironically, as investigated in \cite{EdgeofChaos}, the property of common input synchronization is rather a required property for reservoirs in terms of the reproducibility of the signals (the opposite case is chaotic dynamics).
For robust information processing, the same reservoir is preferred to respond the same according to the identical input stream, even if the initial states of the reservoir differ. 
For the scheme of spatial multiplexing, however, this property acts as a drawback that avoids the duplication of the same reservoir in use.
Accordingly, for spatial multiplexing, preparing a different reservoir or the same setting of the reservoir with different input scaling or with a different choice of qubit for input injections is recommended.

\section{Performance analyses}\label{performance}
In this section, we use numerical experiments to investigate the effect of spatial multiplexing.
By assessing the memory capacity and by using a benchmark task that evaluates the information processing capability to emulate nonlinear dynamical systems called {\it nonlinear auto-regressive moving average} (NARMA) systems, we demonstrate how the order of spatial multiplexing affects the performance of our QR system systematically.
These evaluation schemes adopted here are popular in the context of recurrent neural network learning. 

For the dynamics of QR system, we employ the simplest quantum system, a fully connected transverse-field Ising model, as an example:
\begin{eqnarray} 
H = \sum _{ij} J_{ij} X_i X_j + h Z_i,
\label{eq_Ising}
\end{eqnarray}
where the coupling strengths are randomly assigned such that $J_{ij}$ is distributed randomly from $-J/2$ to $J/2$.
Furthermore, a scale factor $\Delta$ is introduced to make $\tau \Delta$ and $J/\Delta$ dimensionless.
In our numerical experiments, quantum dynamics of the above Hamiltonian is exactly calculated without employing any approximation.

Here, the spatial multiplexing is implemented using QR systems having the same number of qubits $N_{c}$, the input interval $\tau_{c}$, and the virtual nodes $V_{c}$ (which we simply denote $N$, $\tau$, and $V$, from now on) but with different random coupling strengths of $J_{ij}$.
In the following, we see the case when the number of qubits $N$ of a single QR system, which implies the case without spatial multiplexing, is set to $5$.
As an example, we demonstrate in detail when the parameter $\tau \Delta$ is set to 1 and 2 for the memory capacity analyses and for the NARMA tasks, respectively.
We varied the number of the virtual nodes $V$ as 1, 5, and 25 and checked the dependence on the performance.
(Note that the analyses for the different parameter settings, such as the cases for $N=3, 4$ and $\tau \Delta =0.5, 1, 2, 3, 4, 8, 16,$ and $32$, are given in the Appendix.)
Throughout the following experiments, the input stream is randomly drawn from the range $[0, 1]$ and injected to the 1st qubit of each QR system. 
The order of spatial multiplexing, which is defined as the number of QR system driven by a common input stream in parallel, is varied from 1 (without spatial multiplexing) to 5 for the analyses.

\subsection{Memory Capacity}\label{MC}
\begin{figure}
\centering
\includegraphics[width=80mm]{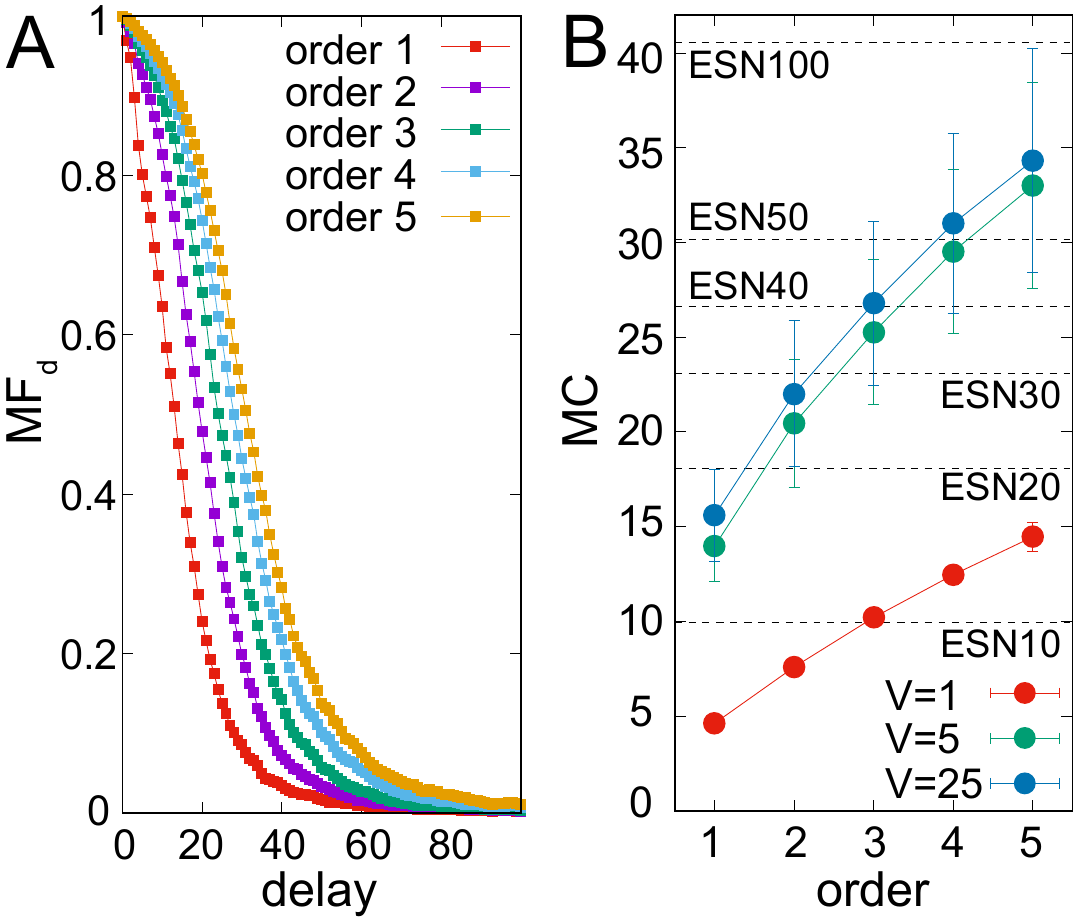}
\caption{{\bf Dependence of the order of spatial multiplexing on memory functions according to the delay and on the memory capacity.}
{\bf A.} The averaged memory functions $MF_d$ are shown with the order of spatial multiplexing varied from 1 to 5. 
Each plot is calculated using 100 trials of different runs with different QR systems, where the virtual nodes are set to 25.
{\bf B.} The averaged MCs are shown according to the order of spatial multiplexing.
Each plot is calculated using 100 trials with different QR systems, and the cases with virtual nodes set to 1, 5, and 25 are overlaid.
As a reference, each plot contains the performance of the conventional ESN. 
The notation ``ESN20,'' for example, represents the averaged MC of the ESN with 20 nodes.
For all the plots, the error bars show the standard deviations.
A single QR system has 5 qubits, and the parameter $\tau \Delta$ is fixed to 1 throughout this analysis. 
}
\label{fig3_new}
\end{figure}

As discussed earlier, the information processing capability of reservoir dynamics can be characterized by its property of transforming the input stream. 
In particular, one of the important characteristics for the computational systems in solving a temporal machine learning task is short-term memory, which is a property to store information of recent inputs to the system's current states.  
Focusing on this point, a measure to evaluate the system's short-term memory property, which is called memory capacity \cite{Jaeger3}, is commonly used.
In this section, we aim to analyze the memory capacity of the QR system and to quantify the effect of the spatial multiplexing in terms of it.
To calculate the measure, the computational system should first learn to reproduce the injected random input of $d$ timesteps before using the current states of the system.
This is equivalent to setting the target output as $\hat{y}_{k} = u_{k-d}$, where $u_{k-d}$ is set as a random sequence ranged in $[0, 1]$ in this study. 

To evaluate the system's emulatability of the target sequence, the memory function $MF_{d}$ is defined as
\begin{equation}
MF_{d} = \frac{cov^{2} (y_{k}, \hat{y}_{k})}{\sigma^{2}(y_{k}) \sigma^{2}(\hat{y}_{k})},
\end{equation}
where $cov(x, y)$ and $\sigma(x)$ express the covariance between $x$ and $y$ and the standard deviation of $x$, respectively.
This measure can take the value from 0 to 1, and as the value gets larger, it suggests that the system's capability to reconstruct the previous input $u_{k-d}$ gets higher.
The memory capacity $MC$ is defined as follows:
\begin{equation}
MC=\sum_{d=0}^{150} MF_{d}.
\end{equation}
As explained in the earlier section, the training scheme of our QR system is based on supervised learning, and for each setting of $d$, the experimental trial consists of a washout phase (2,000 timesteps), a training phase (2,000 timesteps), and an evaluation phase (2,000 timesteps).
Using the time series data of 2,000 timesteps in the training phase and the linear regression explained in Section \ref{learning}, we optimize the readout weights, which we use to calculate the corresponding system output in the evaluation phase.
For each order of spatial multiplexing, we iterated the above procedure by using new QR systems with different random coupling strengths for 100 trials and obtained the averaged $MF_{d}$ and $MC$.

Figure \ref{fig3_new}A shows the averaged $MF_{d}$ over the input delay $d$.
By observing the behavior of $MF_{d}$ against delay $d$, we can see that, according to the increase of the order of spatial multiplexing, the performance gradually improves, showing the relatively large value of $MF_{d}$ in the region of the larger delay.
This tendency can be captured more clearly in the behavior of $MC$ (Fig. \ref{fig3_new}B).
Figure \ref{fig3_new}B plots how the order of spatial multiplexing affects the memory capacity of the QR system in each setting of virtual nodes. 
We can observe that, in all cases, the increase of the order of spatial multiplexing induces the improvement of memory capacity. 
Even in other parameter regions (e.g., for different settings of the parameter $\tau \Delta$ and the number of qubit) of the system, the improved memory capacity, according to the increased order of spatial multiplexing, was generally observed  (see Fig. \ref{figA1_new} in Appendix \ref{A1} for details). 
Interestingly, according to the setting of the parameter $\tau \Delta$, the amount of memory capacity that can be induced was different (Fig. \ref{figA1_new} in Appendix \ref{A1}). 
That is, the memory capacity reaches relatively larger values when $\tau \Delta=0.5$ and $1$ than the other settings of $\tau \Delta$.

\subsection{NARMA tasks}\label{NARMA}
\begin{figure}
\centering
\includegraphics[width=90mm]{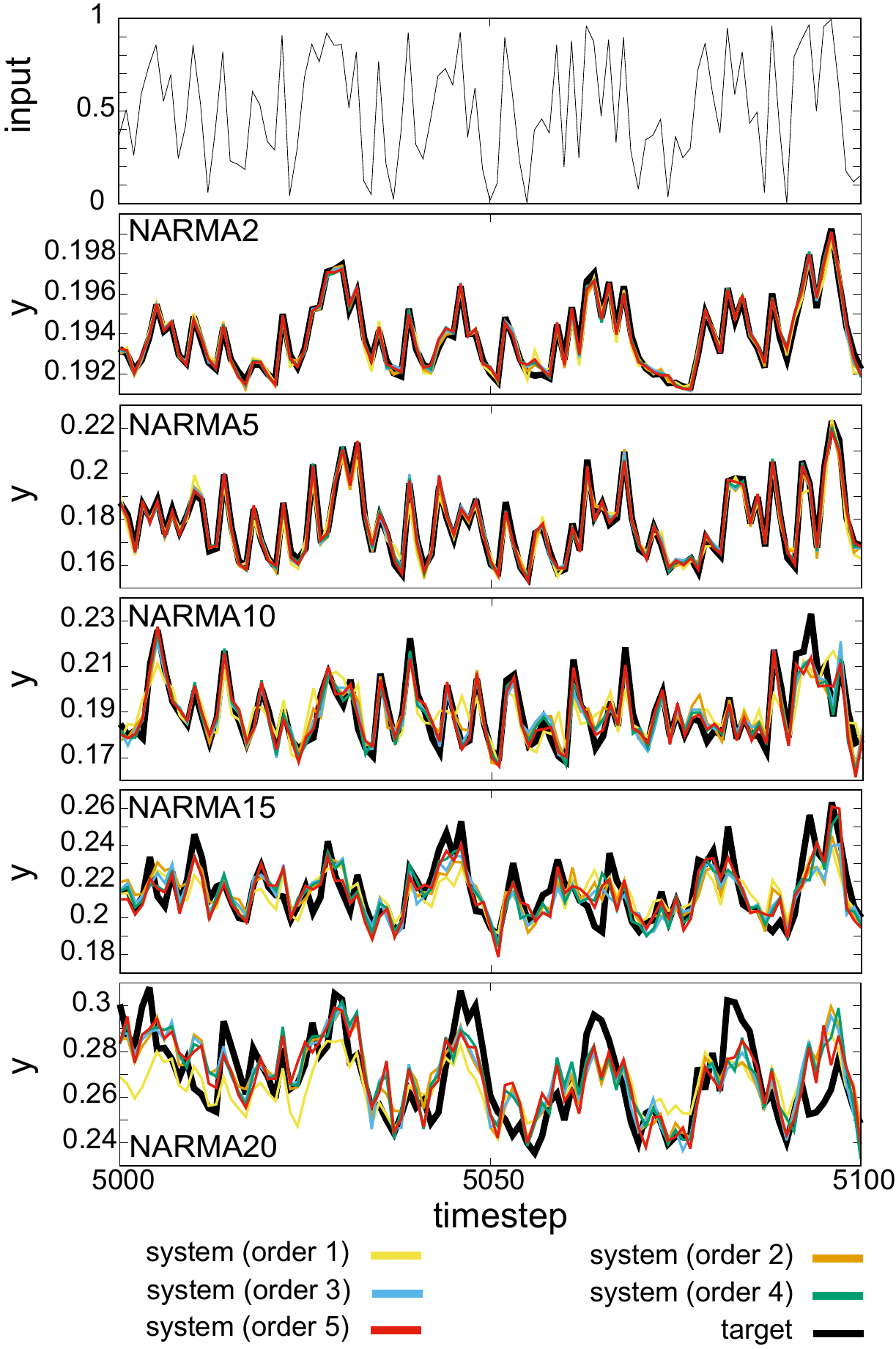}
\caption{{\bf Typical system output time series for the NARMA tasks during the evaluation phase according to the order of spatial multiplexing.}
The uppermost plot shows the random input sequence, and the lower plots show the corresponding task performances for NARMA2, 5, 10, 15, and 20 in order from top to bottom.
Each plot overlays the time series of the target output and system outputs, which exhibit multiplexing until 5 quantum systems with the number of qubits set to 5, the number of virtual nodes set to 25, and the parameter $\tau \Delta$ set to 2.  
}
\label{fig4_new}
\end{figure}
\begin{figure*}
\centering
\includegraphics[width=180mm]{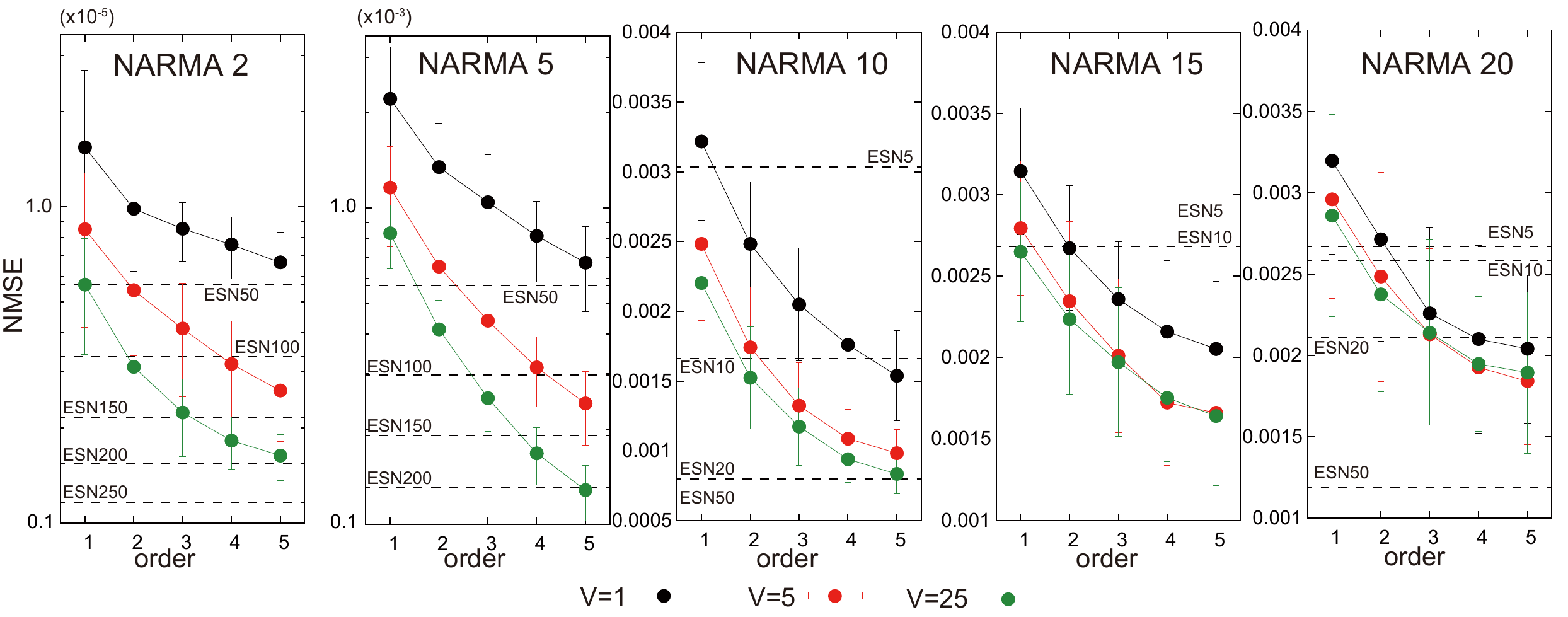}
\caption{{\bf Analysis of the averaged NMSE for the NARMA tasks according to the order of spatial multiplexing.}
For each plot, the cases with virtual nodes set to 1, 5, and 25 are overlaid.
In all the plots, a single QR system has 5 qubits, and the parameter $\tau \Delta$ is fixed to 2. 
The averaged NMSE is calculated using 100 trials with different QR systems.
Note that the y-axis for the plots of NARMA2 and NARMA5 tasks are in the logarithm scale.
As a reference, each plot contains the performance of the conventional ESN. 
The notation ``ESN20,'' for example, represents the averaged NMSE of the ESN with 20 nodes.
For all the plots, the error bars show the standard deviations.
See text for details on the experimental conditions.
 }
\label{fig5_new}
\end{figure*}

The NARMA task is a commonly used benchmark task for evaluating the computational capability of the learning system to implement nonlinear processing with long time dependence.
By calculating the deviations from the target trajectory in terms of errors, the NARMA task tests how well the target NARMA systems can be emulated by the learning system.
According to the choice of the target NARMA system, it is possible to investigate which type of information processing can be performed in the learning system to be evaluated.     
The first NARMA system that we introduce is a second-order nonlinear dynamical system, which was used in \cite{benchmark}, expressed as follows:
\begin{equation}
y_{k} = 0.4y_{k-1} + 0.4y_{k-1}y_{k-2} + 0.6u^{3}_{k} + 0.1.
\end{equation}
We call this system NARMA2 in this paper.
The next NARMA system is the $n$th-order nonlinear dynamical system, which is written as follows:
\begin{equation}
y_{k} = \alpha y_{k-1} + \beta y_{k-1} (\sum_{j=0}^{n-1}y_{k-j-1}) + \gamma u_{k-n+1}u_{k} + \delta,
\end{equation}
where $\alpha, \beta, \gamma,$ and  $\delta$ are $0.3, 0.05, 1.5,$ and $0.1$, respectively.
Here, $n$ varies as $5, 10, 15,$ and $20$, and the corresponding systems are called NARMA5, NARMA10, NARMA15, and NARMA20, respectively.
In particular, NARMA10 is frequently used in the context of evaluating the learning capability of recurrent neural networks (e.g., \cite{benchmark,Reservoir}).
Here, we adopt the multitasking scheme, where the system should simultaneously emulate all the NARMA systems according to the input stream.
For the input stream to the NARMA systems, the range is linearly scaled from $[0, 1]$ to $[0, 0.2]$ to set the range of $y_{k}$ into the stable range.

The learning scheme of our QR system is exactly the same as explained in the previous MC analysis.
Each experimental trial consists of a washout phase (2,000 timesteps), a training phase (2,000 timesteps), and an evaluation phase (2,000 timesteps).
We evaluate the performance by comparing the system output with the target output, which is the {\it normalized mean squared error} (NMSE), expressed as follows:
\begin{equation}
NMSE = \frac{\sum_{k=4001}^{6000} (\hat{y}_{k} - y_{k})^{2}}{\sum_{k=4001}^{6000} \hat{y}_{k}^{2}},
\end{equation}
where $\hat{y}_{k}$ and $y_{k}$ are the target output and the system output at timestep $k$, respectively.
For each $\tau$ setting, NMSEs for all the trials are calculated and averaged for the analysis.
For each order of spatial multiplexing, we iterated the above procedure by using new QR systems with different random coupling strengths for 100 trials and obtained the averaged NMSE.

Figure \ref{fig4_new} shows the typical output time series for the NARMA tasks in the evaluation phase. 
First, it is clearly observed that according to the increase in the order of the NARMA system, the overall task performance gradually worsens, reflecting an increase of the difficulty of the tasks. 
For each NARMA task, according to the increase of the order of spatial multiplexing, we can see that the traceability of the QR system is improved (we can visually confirm this especially for the NARMA5, NARMA10, and NARMA15 tasks in Fig. \ref{fig4_new}). 
These observations can be quantitatively confirmed in the analyses of the averaged NMSE in Fig. \ref{fig5_new}. 
For each setting of the number of virtual nodes (V=1, 5, and 25), the figure plots how the averaged NMSE behaves according to the increase of the order of spatial multiplexing in each NARMA task. 
Figure \ref{fig5_new} shows that for all the NARMA tasks, the increase of the order of spatial multiplexing induces improvements in the task performance. 
In particular, when the order of the NARMA system is 2, 5, and 10, the effect of the increase of the order of spatial multiplexing is significantly high. 
We have checked that this tendency of the effect generally holds for other parameter settings of the QR system (see Fig. \ref{figA2_new} in Appendix \ref{A1} for details).
Furthermore, we have found that, for each NARMA task, a different setting of $\tau \Delta$ exists that shows the best performance through spatial multiplexing. 
For example, in the case for the NARMA2 task and NARMA5 task, the averaged NMSE shows the minimum value when $\tau \Delta = 32$, while in the case for the NARMA15 task and NARMA20 task, $\tau \Delta = 1$ shows the minimum, both through spatial multiplexing of order 5 (Fig. \ref{figA2_new} in Appendix \ref{A1}). 
These findings imply that the parameter $\tau \Delta$ can regulate which type of task the QR system is good at.

\subsection{Temporal versus spatial multiplexing}
\begin{figure*}
\centering
\includegraphics[width=180mm]{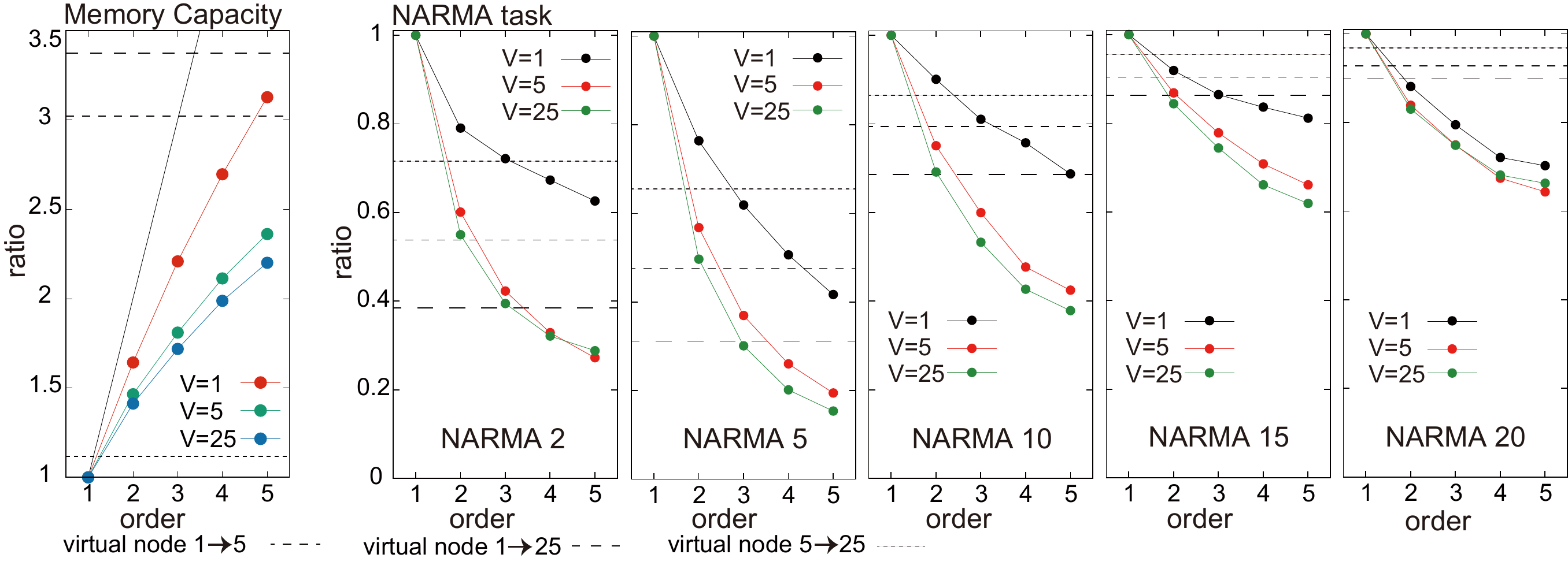}
\caption{{\bf Analyses of the effect of the spatial multiplexing based on the improvement ratio and comparisons with that of the temporal multiplexing.}
Improvement ratios according to the order of spatial multiplexing in terms of the averaged memory capacity (left) and the averaged NMSE for the NARMA tasks (right) are investigated (the parameter settings are the same with the analyses in Sections \ref{MC} and \ref{NARMA}).
In the plot for the memory capacity, the solid line expresses $y=x$ as a reference.
In each plot, the improvement ratios for the temporal multiplexing when the number of virtual nodes is increased from 1 to 5, from 1 to 25, and from 5 to 25 (without spatial multiplexing) are overlaid as a comparison.
For all the analyses, the averaged MC and NMSE used to calculate the improvement ratio in each condition are obtained from the results of 100 trials with different QR systems.
}
\label{fig6_new}
\end{figure*}

\begin{figure}
\centering
\includegraphics[width=80mm]{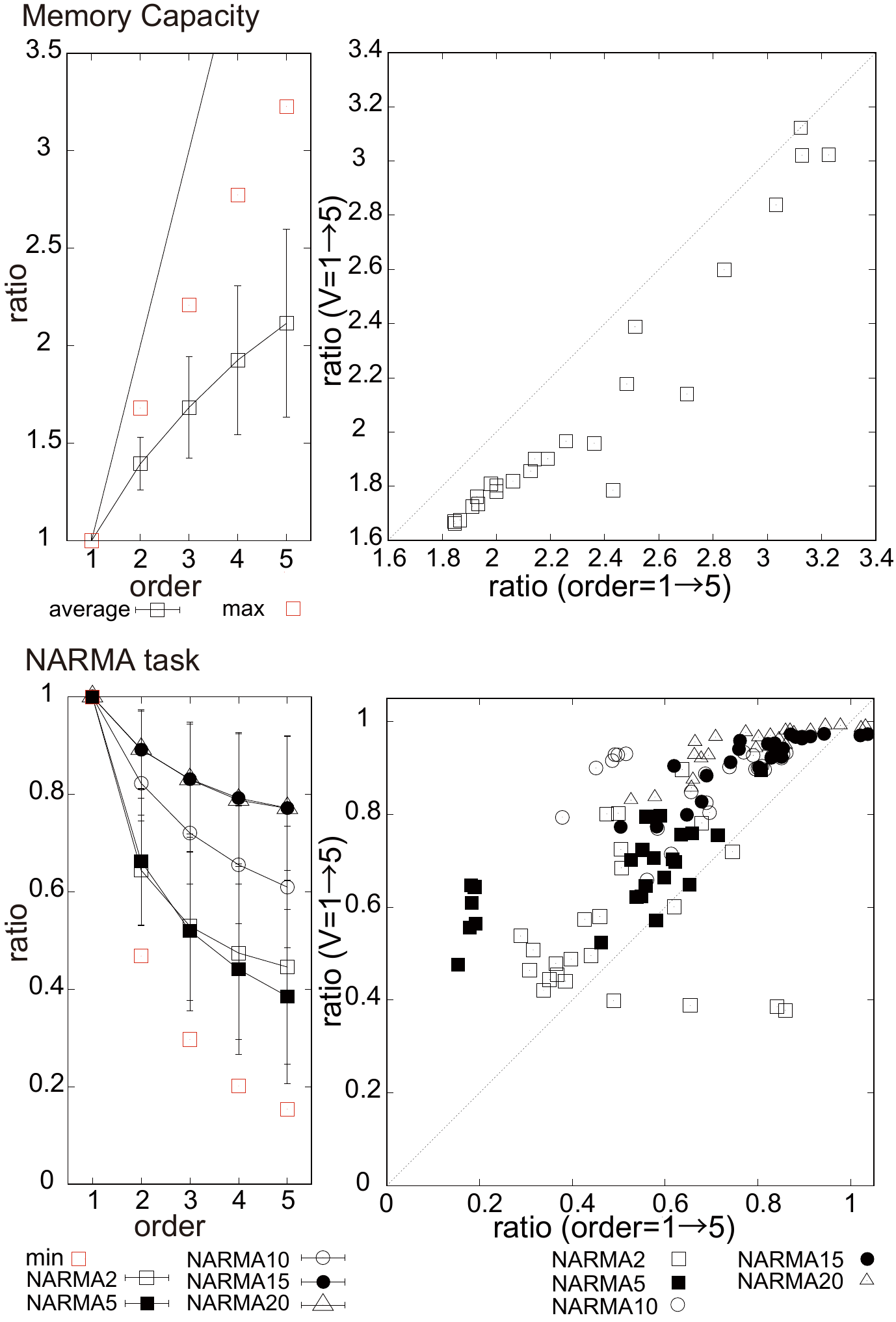}
\caption{{\bf Analyses of the averaged improvement ratios in terms of the averaged memory capacity (upper row) and the averaged NMSE for the NARMA tasks (lower row).} 
According to the order of spatial multiplexing, the improvement ratios are calculated in all the combinations of a number of qubit (3, 4, and 5), virtual node (1, 5, and 25), and $\tau \Delta$ (0.5, 1, 2, 3, 4, 8, 16, and 32), and they are averaged using all these combinations (left diagram in each row).
Note that the solid line in the upper left diagram expresses $y=x$ as a reference.
The error bars show standard deviations, and the maximum or minimum improvement ratio among all the experimental conditions in each order of spatial multiplexing is also plotted. 
The right diagram in each row shows the averaged improvement ratios for spatial multiplexing against those for temporal multiplexing in each analysis.
The ratio is calculated when the order is increased from 1 to 5 with virtual node fixed to 1 for the spatial multiplexing (x axis) and when the virtual node is increased from 1 to 5 without spatial multiplexing for the temporal multiplexing (y axis) to make the increased number of the computational nodes the same for comparisons.
The plots for all the experimental conditions are overlaid.
For all the analyses, the averaged MC and NMSE used to calculate the improvement ratio in each condition are obtained from the results of 100 trials with different QR systems.
}
\label{fig7_new}
\end{figure}

Sections \ref{MC} and \ref{NARMA} have demonstrated that, as the order of spatial multiplexing increases, the memory capacities increase and the performance of the NARMA tasks improves. 
In this section, we analyze the extent to which the order of spatial multiplexing plays a part in these improvements quantitatively. 

Figure \ref{fig6_new} plots how the improvement ratio behaves according to the increase of spatial multiplexing in each experimental case.
The improvement ratio is defined by setting the performance (in terms of the averaged NMSE or MCs) when the order of spatial multiplexing is set to 1 as a basis. 
For both analyses, it is calculated by dividing each averaged MC and NMSE by those when the order of the spatial multiplexing is 1 in each parameter setting, respectively.
As a comparison, the improvement ratios when the number of virtual nodes is increased from 1 to 5, from 1 to 25, and from 5 to 25, without spatial multiplexing (reflecting the effect of temporal multiplexing only), are shown in each plot of Fig. \ref{fig6_new}.
We can clearly observe that in almost all cases, the increase of the order of spatial multiplexing induces the improvements of the performance.

For the memory capacity, we can observe remarkable improvements, where the improvement ratio marks more than twice in all of the setting of virtual nodes, when increasing the order of the spatial multiplexing from 1 to 5 (Fig. \ref{fig6_new}, left diagram).
In particular, the effect of increasing the order of spatial multiplexing from 1 to 5 with the virtual node fixed to 1 ($ratio (order= 1 \rightarrow 5)$) is similar or slightly superior to that of increasing the number of virtual node from 1 to 5 without spatial multiplexing ($ratio (V= 1 \rightarrow 5)$) in terms of capacity (Fig. \ref{fig6_new}, left diagram).
These features were commonly observed in all the experimented parameter settings in this study.
For example, when the order of spatial multiplexing was varied from 1 to 5, the averaged improvement ratio of the memory capacity calculated using all the parameter settings was 2.11, and the maximum improvement ratio among all was 3.23 when $\tau \Delta=0.5$, $V=1$, and $N=5$ (Fig. \ref{fig7_new}, upper left diagram).
Furthermore, we observed ``$ratio (order= 1 \rightarrow 5) > ratio (V= 1 \rightarrow 5)$'' in almost all the parameter settings (Fig. \ref{fig7_new}, upper right diagram), which characterizes the range of effectiveness of the spatial multiplexing.

For the NARMA task, by increasing the order of spatial multiplexing, the value of the improvement ratio is decreased, suggesting the improvements of the performance (Fig. \ref{fig6_new}, right diagrams). 
In particular, in the NARMA5 task when $V=25$, the value decreased by a factor of 10 when the order of spatial multiplexing was varied from 1 to 5. 
Interestingly, this improvement ratio is much superior to that of increasing the virtual node from 1 to 25 despite the larger increase in the computational nodes, which implies that cases exist in which the increase of the order of spatial multiplexing behaves superior to that of temporal multiplexing.
These tendencies follow in all the experimented parameter settings in this study.
The performance of each NARMA task improves in each parameter setting by increasing the order of spatial multiplexing (Fig. \ref{fig7_new}, lower left diagram).
For example, when the order of spatial multiplexing was varied from 1 to 5, the averaged improvement ratio calculated using all the parameter settings was 0.39 in the NARMA5 task, and the minimum improvement ratio among all was 0.15 when $\tau \Delta=2$, $V=25$, and $N=5$ in the NARMA5 task (Fig. \ref{fig7_new}, lower left diagram).
Similarly to the case for the memory capacity, in each NARMA task, we observed ``$ratio (order= 1 \rightarrow 5) < ratio (V= 1 \rightarrow 5)$'' in almost all the parameter settings (Fig. \ref{fig7_new}, lower right diagram).
These results suggest that in some cases, spatial multiplexing adds a more effective number of computational nodes than does temporal multiplexing.

\section{Toward engineering quantum reservoir through spatial multiplexing}
As we saw in Section \ref{theory} and demonstrated in \ref{performance}, spatial multiplexing improves the performance of the system.
In this section, we provide a few notes on the possibility to engineer QR through the spatial multiplexing scheme.
As we discussed in Section \ref{theory}, although we can improve performance by increasing the order of spatial multiplexing in theory, this does not always apply in actual experiments because of overfitting.
In such cases, limiting the number of computational nodes is preferable.
Given a fixed number of computational nodes, we investigate in this section a method to engineer the efficient combinations of reservoirs.

Similar to Section \ref{theory}, let us assume that we have three reservoirs, A, B, and C, with reservoir A having $N$ computational nodes, and reservoirs B and C having the same number of nodes $N'$.
We also assume that these reservoirs satisfy the basic properties of the regression equation setting and least squares solutions presented in Section \ref{theory}.
At first, the inequality $r_{A+B+C}^{2}\leq min\{ r_{A}^{2}, r_{B}^{2}, r_{C}^{2}, r_{A+B}^{2}, r_{B+C}^{2}, r_{A+C}^{2} \}$ suggests that combining reservoirs A, B, and C performs best if we could avoid overfitting in practice.
Now, by retaining the total number of nodes fixed to $N+N'$, we determine the better choice between reservoir B or C for combination with reservoir A to improve performance. 

At first glance, choosing the reservoir that has better performance is preferable.
However, this is not always the case.
Given that reservoir B has better performance than reservoir C, that is, $r_{B}^{2} \leq r_{C}^{2}$, $r_{A+B}^{2} \leq r_{A}^{2}$ and $r_{A+C}^{2} \leq r_{A}^{2}$ hold, but $r_{A+B}^{2} \leq r_{A+C}^{2}$ does not hold in general.
(We can easily find a counter example such as $y= \begin{pmatrix} 1 & 1 & 1 \end{pmatrix}^{\mathrm{T}}$, $X_{A}= \begin{pmatrix} 0.25 & 1 & 0 \end{pmatrix}^{\mathrm{T}}$, $X_{B}= \begin{pmatrix} 1 & 0 & 0 \end{pmatrix}^{\mathrm{T}}$, $X_{C}= \begin{pmatrix} 0 & 1 & -1 \end{pmatrix}^{\mathrm{T}}$.)
We then apply the relations we obtained in Section \ref{theory} to reservoir A+B and A+C, which are 
\begin{align*}
&\max\{(r_{A}^{2}-\lambda_{Q_{A, A+B}} ||y||^{2}), (r_{B}^{2}-\lambda_{Q_{B, A+B}} ||y||^{2}) \} \\
&\leq r_{A+B}^{2} \leq \min\{ r_{A}^{2}, r_{B}^{2} \},
\end{align*}
and
\begin{align*}
&\max\{(r_{A}^{2}-\lambda_{Q_{A, A+C}} ||y||^{2}), (r_{C}^{2}-\lambda_{Q_{C, A+C}} ||y||^{2}) \} \\
&\leq r_{A+C}^{2} \leq \min\{ r_{A}^{2}, r_{C}^{2} \}.
\end{align*}
Given $r_{B}^{2} \leq r_{C}^{2}$, to evaluate $r_{A+B}^{2}$ and $r_{A+C}^{2}$, we need to check how these ranges overlap.
Only if no overlap exists, can we safely predict the reservoir to add without actually performing the task.
When $r_{A}^{2} \leq r_{B}^{2} \leq r_{C}^{2}$, these two ranges always overlap because $r_{A+B}^{2} \leq r_{A}^{2}$ and $r_{A+C}^{2} \leq r_{A}^{2}$.
When $r_{B}^{2} \leq r_{A}^{2} \leq r_{C}^{2}$ or $r_{B}^{2} \leq r_{C}^{2} \leq r_{A}^{2}$, and if $r_{B}^{2} < \max\{(r_{A}^{2}-\lambda_{Q_{A, A+C}} ||y||^{2}), (r_{C}^{2}-\lambda_{Q_{C, A+C}} ||y||^{2}) \}$ holds, then we can safely decide to choose reservoir B as the appropriate partner for combination without actually performing the task, because it satisfies $r_{B}^{2} < r_{A+C}^{2}$ and accordingly $r_{A+B}^{2} < r_{A+C}^{2}$ holds.

Furthermore, although we demonstrated spatial multiplexing by combining QR systems that have different coupling strengths with the other parameters fixed in our numerical experiments, note that the combination can consist of any reservoirs if they are not synchronized.
The choice of the combination depends on the efficiency of the usage in each experimental setting.
For example, parameter $\tau \Delta$ is dependent on the energy applied to the experimental platform and can be regulated if we consider energy efficiency.

\section{Discussion}
In this paper, we have introduced a scheme, spatial multiplexing, to boost the computational power in QRC. 
Considering the physical experiment, this scheme is operationally easy to implement but is remarkably effective, and we have theoretically shown that the scheme inevitably increases the computational power. 
The effect was demonstrated through numerical experiments using a number of benchmark tasks, and the performance of learning was observed to be improved. 
We have also examined the theoretical implications of the proposed scheme and discussed its range of validity and limitations, which would be useful and applicable not only for QRC but also for reservoir computing in general, including the case for conventional software implementations.

Although the scheme of spatial multiplexing allows us to efficiently increase the computational nodes, we should be sensitive to the case of overfitting in practical applications. 
In our experiments, we observed several performances, which were thought to be caused by overfitting (e.g., the results of NARMA tasks in higher values of $\tau \Delta$ (Fig. \ref{figA2_new} in Appendix \ref{A1})). 
To avoid these situations, one can introduce a Ridge regression or Lasso for the training procedure, which assigns a penalty to readout weights for regressions. 
By combining with these sparse regressions, one can establish a scheme to selectively exploit effective degrees of freedom from massive computational nodes increased by spatial multiplexing.

NMR ensemble system has been regarded as a strong candidate for physical platform of QRC.
In NMR quantum reservoir system, the spatial multiplexing with some different molecules, introduced in Sec.\ref{sec:SM}, is an easier option to increase the computational power than increasing the number of addressable qubits.
Another option is increasing the number of unaddressable qubits and virtual nodes, which will be introduced with a detail in our future work. 
We can also introduce an easier implementation of spatial multiplexing even with the same molecule with NMR pulse techniques to change the interaction Hamiltonian effectively \cite{Negoro3,Negoro4}.
The pulse techniques are often utilized for quantum simulation experiments.
For example, Ising type Hamiltonian $X_i X_j$ can be changed to $X_i X_j + Y_i Y_j + a Z_i Z_j$ for any parameters $a$, with applying the multiple pulse sequence with a parameter for spacing between pulses, $a$ \cite{Negoro5}.
It was shown in a quantum simulation experiment \cite{Negoro6} that the dynamical behavior of a nuclear spin system with the interactions $X_i X_j + (2a-1) Y_i Y_j - 2a Z_i Z_j$ are substantially different depending on $a$.
Just with changing the parameter of applying pulse, we can easily implement the spatial multiplexing with some different quantum dynamical systems in the same molecule.

Spatial multiplexing will offer an opportunity to increase the computational nodes and boost the computational power not only for QRs but also for any interacting systems that contain components that are operationally or experimentally difficult to manipulate and increase. 
By extending this line of thought, we can develop a concept of composing multiple reservoirs, each with different physical systems. 
For example, it might be worth composing photonic and quantum systems and treating them as one entire reservoir in some applications. 
According to how this scheme is applied in the real world, this concept would create options from which to flexibly choose the physical systems to use as a computational resource in a given situation.

Finally, one of the intriguing flavors in the framework of QRC is its exploitation of the quantum computational supremacy region, where the system possesses exponential degrees of freedom as hidden nodes. 
We reiterate that, as spatial multiplexing increases true nodes proportionally to its order, its increase of hidden nodes is also proportional, while increasing the number of qubits in the interacting system will directly lead to exponential increase in hidden nodes. 
This fact implies that, even if we have the same number of true nodes, the number of hidden nodes can differ according to how the spin-ensemble molecular samples were prepared; hence, the computational power and preference would also differ. 
We suggest each experimenter to regulate how to prepare their reservoirs based on their given experimental conditions and their operability of the system, and we believe that the spatial multiplexing technique will become one of the common and practical options for boosting the computational power of QRs in the near future.

\section{Acknowledgements}
K.N. would like to acknowledge Taichi Haruna for fruitful discussions and Hiroki Masui for his help in arranging figures.
K.N., K.F., and M.N. are supported by JST PRESTO Grant Number JPMJPR15E7, JPMJPR1668, and JPMJPR1666, Japan.
K.M is also supported by JST PRESTO Grant Number JPMJPR1666, Japan.
K.N. is supported by KAKENHI No. 16KT0019, No. 15K16076, and No. 26880010. 
K.F. is supported by KAKENHI No.16H02211, JST ERATO Grant Numnber JPMJER1601, and JST CREST Grant Number JPMJCR1673.

\section{Author Contributions}
K.N. and K.F. designed and conducted numerical experiments.
All authors discussed the results and implications and wrote the manuscript at all stages.

\section{Author Information}
There is no competing financial interests.

\appendix
\section{Extended numerical experiments and analyses}\label{A1}
\begin{figure*}
\centering
\includegraphics[width=180mm]{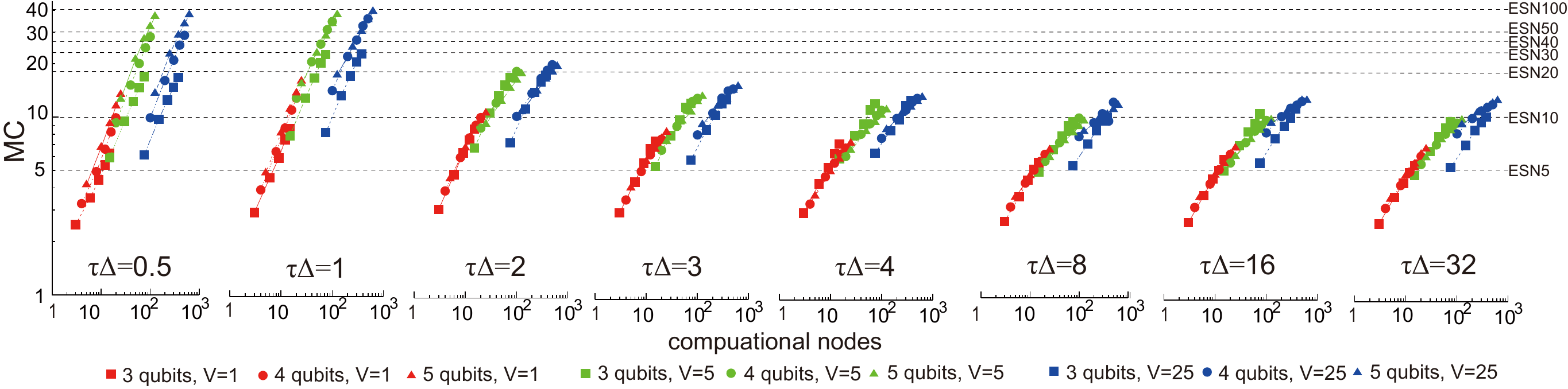}
\caption{{\bf Plots showing the effect of the spatial multiplexing in terms of the averaged MC for each number of qubits of a single QR system, each number of virtual nodes, and each parameter $\tau \Delta$.}
For each plot, the horizontal axis represents total computational nodes in a QR system, and the vertical axis represents the averaged MC.
The parameter $\tau \Delta$ is varied as $0.5, 1, 2, 3, 4, 8, 16,$ and $32$.
The number of qubits in a single QR system is represented by different point shapes (square, circle, and triangle indicate the cases for 3, 4, and 5 qubits, respectively).
The number of virtual nodes is represented as a difference in color (the number of virtual nodes $V=1, 5,$ and $25$ are represented as black, red, and green, respectively).
The plots connected with lines represent the results when the order of spatial multiplexing is increased from 1 to 5 (this can be seen from the increase in the total number of computational nodes) with other system parameters fixed.
As a reference, each plot contains the performance of the conventional ESN. 
The notation ``ESN20,'' for example, represents the averaged MC of the ESN with 20 nodes.}
\label{figA1_new}
\end{figure*}

\begin{figure*}
\centering
\includegraphics[width=170mm]{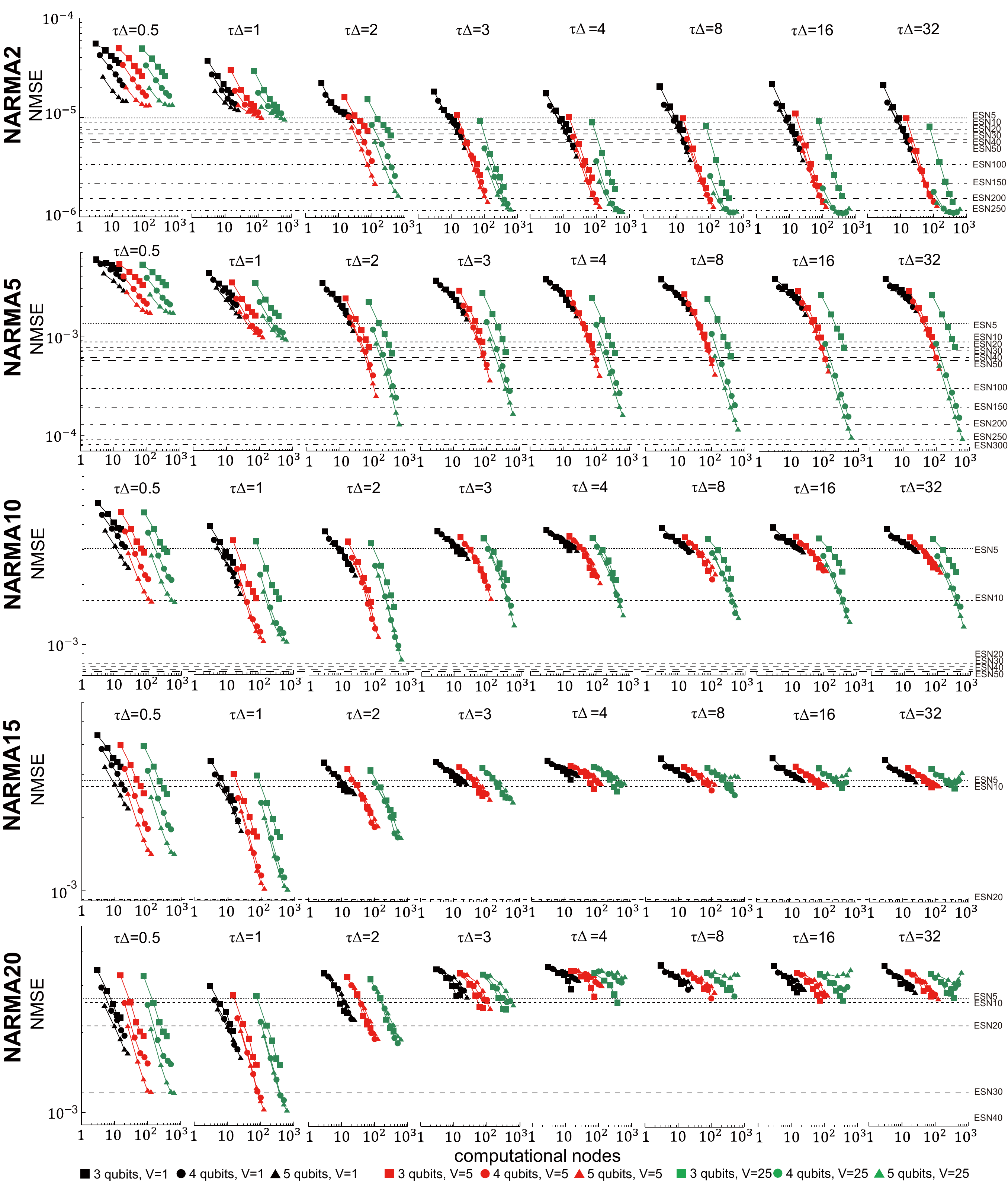}
\caption{{\bf Plots showing the effect of the spatial multiplexing in terms of the averaged NMSE for the NARMA tasks for each number of qubits of a single QR system, each number of virtual nodes, and each parameter $\tau \Delta$.}
The cases for NARMA2, 5, 10, 15, and 20 are investigated. 
For each plot, the horizontal axis represents the total computational nodes in a QR system, and the vertical axis represents the averaged NMSE.
The parameter $\tau \Delta$ is varied as $0.5, 1, 2, 3, 4, 8, 16,$ and $32$.
The number of qubits in a single QR system is represented by different point shapes (square, circle, and triangle indicate the cases for 3, 4, and 5 qubits, respectively).
The number of virtual nodes is represented as a difference in color (the number of virtual nodes $V=1, 5,$ and $25$ are represented as black, red, and green, respectively).
The plots connected with lines represent the results when the order of spatial multiplexing is increased from 1 to 5 (this can be seen from the increase in the total number of computational nodes) with other system parameters fixed.
As a reference, each plot contains the performance of the conventional ESN. 
The notation ``ESN20,'' for example, represents the averaged NMSE of the ESN with 20 nodes.
}
\label{figA2_new}
\end{figure*}

In the main text, we showed the results of the numerical experiments for the QR system with its system parameters set to $N=5$ and $\tau \Delta=1, 2$.
In this section, we show thorough and systematic analyses of different parameter settings, varying $N$ as 3, 4, and 5 and varying $\tau \Delta$ as $0.5, 1, 2, 3, 4, 8, 16,$ and $32$, which are summarized in Fig. \ref{figA1_new} and Fig. \ref{figA2_new}.

\section{Echo state network settings for comparisons}
To characterize the computational capability of our system, in the main text, we compared its performance of the NARMA tasks and its memory capacity with those of a conventional ESN \cite{Jaeger0,Jaeger1,Jaeger0SI}.
This section explains in detail the settings of the ESN used for the comparisons.

The ESNs are a type of random recurrent neural network that consists of internal computational nodes (the number of internal computational nodes is denoted as $N_{ESN}$), input nodes, and output nodes. 
The activation of the $i$th internal node at timestep $k$ is expressed as $x^{i}_{k}$.
The weights $w^{ij}$ for the internal network connect the $i$th node to the $j$th node, and the input weights $w^{i}_{in}$ connect the input node to the $i$th internal node.
Internal computational nodes with one bias are connected to the output unit through readout weights $w^{i}_{out}$, where $x^{0}_{k} = 1$ and $w^{0}_{out}$ is assigned for the bias term. 
Learning of the readout weights $w^{i}_{out}$ is performed using the same procedure explained in the main text for each task.
The internal weights $w^{ij}$ are randomly determined from the range $[-1.0, 1.0]$, and the spectral radius of the weights is regulated according to the setting for each task, as explained below.
Similarly, the input weights $w^{i}_{in}$ are randomly determined from the range $[-\sigma, \sigma]$, where $\sigma$ is a scaling parameter explained later.
The time evolution of the ESN is expressed as follows:
\begin{align}
x^{i}_{k} &= f(\sum^{N_{ESN}}_{j=1}w^{ij}x^{j}_{k-1} + w^{i}_{in}u_{k}), \\
y_{k} &= \sum^{N_{ESN}}_{i=0}w^{i}_{out}x^{i}_{k},
\end{align}
where $f(x)$ is set as $\tanh(x)$ in this paper.
To make a fair comparison of the task performance, the I/O setting of the ESN was set to be the same as that of our system for each task. 
For example, the lengths of the washout, training, and evaluation phases and the evaluation procedures were kept the same.
The detailed experimental conditions are given for each of these comparisons below.

For the NARMA task, we first prepared 10 different ESNs for each setting of $N_{ESN}$, which vary as 5, 10, 20, 30, 40, 50, 100, 150, 200, 250, and 300.
The scaling parameter of the input weights $\sigma$ is varied as 1.0, 0.5, 0.2, 0.1, 0.05, 0.01, 0.005, and 0.001, and the spectral radius of the internal weights is also varied from 0.1 to 2.0 in increments of 0.1.
For each ESN, by fixing the spectral radius and the parameter $\sigma$, we ran 10 different trials, driven by different random input sequences, and test the emulation tasks of all the NARMA systems (NARMA2, 5, 10, 15, and 20) using a multitasking scheme for each trial.
After performing all the trials of the NARMA tasks with all the parameter settings varied for each ESN having the computational node $N_{ESN}$, we collected the lowest NMSE, which indicates the best performance in this experiment corresponding to the ESN, and calculated the averaged NMSE for each NARMA task over 10 different ESNs for each setting of $N_{ESN}$.
These averaged NMSEs were used for comparison.   

To evaluate the memory capacities, 100 different ESNs were driven by different random input sequences with a spectral radius fixed at 0.9 and the scaling parameter of the input weights fixed to $\sigma=0.01$. 
The emulation tasks of 5 dynamical systems with different degrees of nonlinearity, which are explained in the main text, were performed for each trial using a multitasking scheme.
Analyses of the performance were conducted using the same procedures used by our system and defined in the main text, and the averaged capacities were calculated using these 100 trials and used for comparison.


\begin{thebibliography}{99}
\bibitem{IBM_Neuro}
P.A. Merolla {\it et al.}, Science {\bf 345}, 668 (2014).

\bibitem{Jaeger0} 
H. Jaeger and H. Haas, Science {\bf 304}, 78 (2004).

\bibitem{Maass0} 
W. Maass, T. Natschl\"{a}ger, and H. Markram, Neural Comput. {\bf 14}, 2531 (2002).

\bibitem{Reservoir} 
D. Verstraeten, B. Schrauwen, M. D'Haene, and D. Stroobandt, Neural Netw. {\bf 20}, 391 (2007).

\bibitem{Neuromorphic0} 
A. Z. Stieg {\it et al.}, Adv. Mater. {\bf 24}, 286 (2012).

\bibitem{Bucket} 
C. Fernando and S. Sojakka, 
{\it Pattern recognition in a bucket}
In Lecture Notes in Computer Science {\bf 2801}, p. 588 (Springer, 2003).

\bibitem{Laser0} 
L. Appeltant {\it et al.}, Nat. Commun. {\bf 2}, 468 (2011).

\bibitem{Laser1} 
L. Larger {\it et al.}, Optics Express {\bf 20}, 3241 (2012).

\bibitem{Spintronics}
J. Torrejon {\it et al.}, Nature {\bf 547}, 428 (2017).

\bibitem{Kohei1} 
K. Nakajima {\it et al.}, Front. Comput. Neurosci. {\bf 7}, 1 (2013).

\bibitem{Kohei2} 
K. Nakajima {\it et al.}, J. R. Soc. Interface {\bf 11}, 20140437 (2014).

\bibitem{Kohei3} 
K. Nakajima {\it et al.}, Sci. Rep. {\bf 5}, 10487 (2015).

\bibitem{Kohei4} 
K. Nakajima {\it et al.}, Soft Robotics (in press).

\bibitem{QR}
K. Fujii and K. Nakajima, Phys. Rev. Applied {\bf 8}, 024030 (2017).

\bibitem{NMRQC1}
D. G. Cory {\it et al.}, Fortschr. Phys. {\bf 48}, 875 (2000). 

\bibitem{NMRQC2}
J. A. Jones, Prog. Nucl. Magn. Reson. Spectros. {\bf 59}, 91 (2011).

\bibitem{TM_note} 
In \cite{QR}, the procedure of temporal multiplexing is called {\it time multiplexing}, but is the same thing.
We used the term ``temporal'' to coordinate with the term ``spatial'' in this paper.

\bibitem{Negoro1}
C. Negrevergne {\it et al.}, Phys. Rev. lett. {\bf 96}, 170501 (2006).

\bibitem{Negoro2}
D. Lu {\it et al.}, npj Quantum Information {\bf 3}, 45 (2017).

\bibitem{EdgeofChaos} 
N. Bertschinger and T. Natschl\"{a}ger, Neural Comput. {\bf 16}, 1413 (2004).

\bibitem{CommonNoise} 
R. Toral, C. R. Mirasso, E. Hern\'{a}ndez-Garcia, and O. Piro, CHAOS {\bf 11}, 665 (2001).

\bibitem{Jaeger3} 
H. Jaeger, GMD Report {\bf 152}, German National Research Center for Information Technology (2001).

\bibitem{benchmark} 
A. F. Atiya and A. G. Parlos, IEEE Trans. Neural Netw. {\bf 11}, 697 (2000).

\bibitem{Negoro3}
U. Haeberlen,
{\it High resolution NMR in solids --- Selective averaging},
in Advances in Magnetic Resonance Suppl. 1 (Academic Press, New York, 1976).

\bibitem{Negoro4}
L. M. Vandersypen and I. L. Chuang, Reviews of Modern Physics {\bf 76}, 1037 (2005).

\bibitem{Negoro5}
G. Roumpos, C. P. Master, and Y. Yamamoto, Phys. Rev. B, {\bf 75}, 094415 (2007).

\bibitem{Negoro6}
G. A. Alvalez, D. Suter, and R. Kaiser, Science {\bf 349}, 846 (2015).

\bibitem{Jaeger1} 
H. Jaeger, GMD Report {\bf 159}, German National Research Center for Information Technology (2002).

\bibitem{Jaeger0SI} 
H. Jaeger, GMD Report {\bf 148}, German National Research Institute for Computer Science (2001).
\end{thebibliography}
\end{document}